\documentclass[12pt,a4paper]{article}
\usepackage{graphicx} 
\DeclareGraphicsExtensions{.pdf,.jpg,.png,.eps,.svg}
\graphicspath{{./Images/}}
\usepackage{siunitx}
\usepackage{subcaption}
\usepackage{wrapfig}
\usepackage[paper=a4paper,margin=2.5cm,top=2cm,bottom=2cm]{geometry}
\usepackage[utf8]{inputenc}
\usepackage[main=english, ngerman]{babel}
\usepackage{csquotes}
\usepackage{authblk}
\usepackage[backend=biber,
maxbibnames=100,
style=nature,
url=false,
isbn=false,
sorting=none
]{biblatex}
\title{\textit{In-situ} compression and shape recovery of Ceramic single grain micro-pillar}
\author{Justin Jetter}
\author{Eckhard Quandt}
\affil{Chair for Inorganic Functional Materials, Kiel University, Germany}
\date{}

\begin{document}
\maketitle

\section{Abstract}
Most ceramic materials are known for high fracture toughness while reacting highly brittle to physical deformation. Some advancements were made by utilizing the transformation toughening effect of Yttria-doped Zirconia. However, finding a ceramic material demonstrating an effect analogous to the Shape Memory Effect (SME) in certain metals, that also allows for superelastic responses, remains a challenge. The underlying mechanism for SME and superelasticity is based on crystallographic variations within the material's grains, requiring sophisticated electron microscopy techniques for direct observation. The combination of a scanning electron microscope (SEM) with focused ion beam (FIB) milling, a \textit{Kleindiek Nanotechnik GmbH} micro-manipulator with a \qty{1.5}{\um} diamond tip, and the ability to achieve \textit{in-situ} heating up to \qty{450}{\celsius} on a \textit{Kleindiek} heating stage provides a robust platform for the preparation, deformation, and heating of micro-pillars made from ceramic materials. This setup enabled us to conduct detailed studies on the Zirconia-based ceramic, observing permanent deformation exceeding \qty{4}{\percent} strain, followed by shape recovery at \qty{370}{\celsius}. The paper provides outlines the key experimental steps that facilitated these observations.

\section{Introduction}
In recent years, extensive effort has been devoted to identifying ceramic materials that can absorb physical deformation and recover their previous dimensions \cite{Lai2013,Zeng2016,Du2015a,Pang2019}. The primary focus has been on tuning the crystallographic lattice of the material to satisfy the compatibility conditions known to lead to high-performance shape memory alloys such as NiTi-based alloys \cite{Ball1987,Song2013,Chluba2015,Gu2018a}. A significant challenge in ceramics is the phase transition between the monoclinic low-temperature and the tetragonal high-temperature phase. This transition is more rigid compared to the cubic transformations seen in shape memory alloys \cite{Otsuka2005}, necessitating the activation of multiple correspondences to enhance reversibility \cite{Gu2021a}.
By reducing the deformed volume to the scale of single grains, it becomes easier to isolate structures that rely on a single correspondence \cite{Zhang2021}. The material investigated in this study has a chemical composition of \qty{14}{\percent}YO$_{1.5}$ + \qty{14}{\percent}TaO$_{2.5}$ + \qty{2}{\percent}ErO$_{1.5}$ + \qty{70}{\percent}ZrO$_2$ and was specifically chosen for its optimized crystallographic compatibility, with a deviation from the ideal middle eigenvalue $\lambda_2=1$ quantified as $1-\lambda_2=0.00349$ based of the lattice parameters calculated from X-ray diffraction (XRD) measurements. The composition was identified using machine learning methods aimed at discovering highly compatible ceramics, details of which will be presented in an upcoming publication \cite{Pandey2025}.

\section{Experimental}
The ceramic bulk sample used for the experiment was fabricated from raw oxide powders mixed in the correct ratio by following a solid-state sintering procedure with at a maximum temperature of \qty{1500}{\celsius}. A detailed description of the fabrication process was published in a prior publication \cite{Jetter2020}. To characterize its thermal properties and confirm the presence of the monoclinic phase at room temperature (RT), differential scanning calorimetry (DSC) and X-ray diffraction (XRD) analyses were conducted. DSC provided transformation temperatures ($A\textsubscript{s}$, $A\textsubscript{f}$, $M\textsubscript{s}$, and $M\textsubscript{f}$), indicating that the composition could exist in either the austenite or martensite phase at RT, depending on its thermal history. XRD measurements were performed on the as-sintered sample, revealing a predominant tetragonal phase, as indicated by the highest intensity peak at $\sim \qty{30}{\degree}$ $2\theta$.
If the conditions $M\textsubscript{s} \qty{<0}{\celsius}$ and $A\textsubscript{s} \qty{>50}{\celsius}$ are satisfied, these temperatures suggested that the RT phase could be changed to monoclinic, the sample was immersed in liquid nitrogen ($T\textsubscript{LN} = \qty{-196}{\celsius}$) to induce this transformation. A subsequent XRD measurement confirmed the transition to the monoclinic phase, serving as the basis for further experiments using scanning electron microscopy (SEM) and focused ion beam (FIB) preparations. Compression experiments were performed in a FEI Helios 600 FIB-SEM, equipped with a \textit{Kleindiek} micro-manipulator MM3A-EM. The micro-manipulator features a flat-top diamond tip with a contact area of \qty{1.5}{\um} in diameter. The sample was mounted onto a miniature heating stage (MHS-450) which itself was positioned onto a spring table (spring constant $=\qty{1158}{\newton\per\m}$) and the corresponding force analysis software is able to use relative displacements within an SEM image to calculate force-displacement curves. Both of these pieces of hardware were also manufactured by \textit{Kleindiek GmbH}.

\begin{wrapfigure}{r}{0.31\textwidth}
\vspace*{-1em}
\centering
\includegraphics{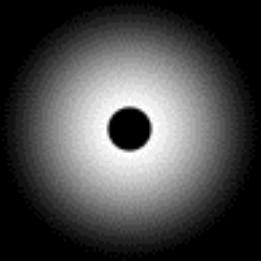}
\caption{Grayscale bitmap for selective FIB cut. Outside diameter: \qty{12}{\um}, Inside diameter: \qty{2}{\um}}
\label{fig:bitmap}
\vspace*{-1em}
\end{wrapfigure}

At suitable magnification, the electron beam was used to identify surface grains with sufficiently large diameter (\qty{>3}{\um}). The ion beam then carved a single micropillar (\qty{\sim 1}{\um} in diameter) by selectively removing surrounding material. A grayscale bitmap mask (Figure~\ref{fig:bitmap}) facilitated material removal in a controlled fashion, creating a crater approximately \qty{12}{\um} in diameter around the pillar. The pillar was then polished using a circular mask that progressively removed material from the outer edge inward, minimizing redeposition artifacts.
A sample tilt of \qty{52}{\degree} toward the secondary electron detector enabled a suitable view of the pillar's shape, allowing precise positioning of the \textit{Kleindiek} \qty{1.5}{\um} diamond flat-top indenter. As both the micromanipulator and the sample were mounted on the same plate, relative movement between them was limited to the controlled actions of the micromanipulator.
The applied load on the micropillar was measured by image tracking of relative displacement of the sample and indenter tip. Since the sample was positioned on a calibrated spring table with a known spring constant, the Kleindiek software is able to calculate the applied force while accounting for magnification and stage tilt. By measuring the diameter ($d=\qty{1.9}{\um}$) and the height of the undeformed pillar ($h=\qty{5.8}{\um}$) those values are converted to stress-strain values.
\textit{In-situ} heating experiments were conducted using a \textit{Kleindiek} heating stage, capable of reaching \qty{450}{\celsius} while maintaining vacuum conditions. Continuous SEM observation of the deformed pillar throughout the heating process allowed for real-time monitoring of shape recovery.

\begin{figure}[t]
    \centering
    \begin{subfigure}[t]{0.47\linewidth}
        \includegraphics[width=\linewidth]{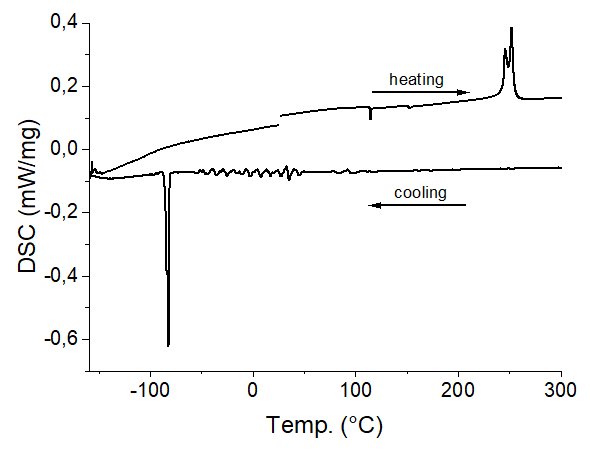}
        \caption{DSC measurement to determine transformation temperatures. Transformation peaks visible at \qty{\sim-82}{\celsius} and \qty{\sim250}{\celsius}.}
        \label{fig:DSC}
    \end{subfigure} 
    \hfill
    \begin{subfigure}[t]{0.47\linewidth}
        \centering
        \includegraphics[width=\linewidth]{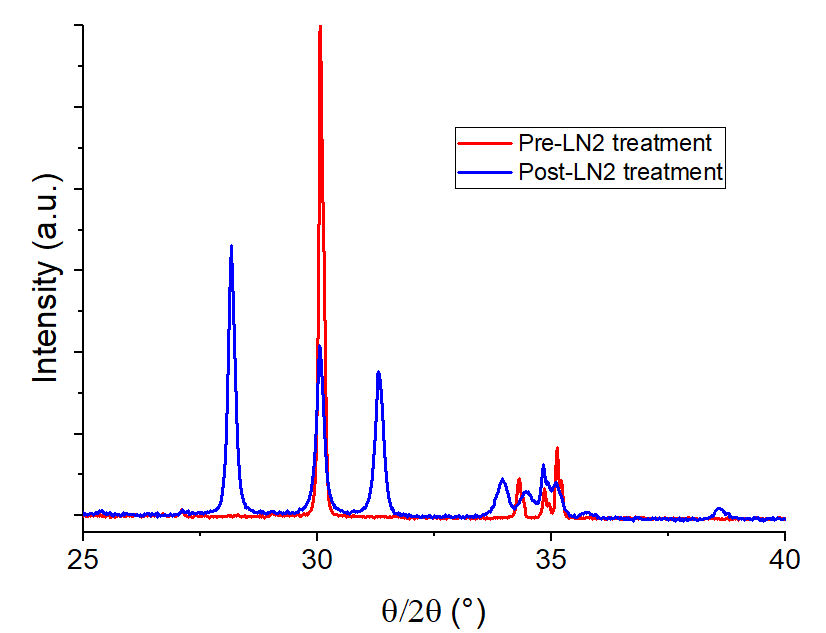}
        \caption{XRD diffraction pattern of the sample before and after LN2 treatment. Tetragonal (red) prior to LN2 treatment and monoclinic (blue) afterwards.}
        \label{fig:XRD}
    \end{subfigure}
    \caption{DSC and XRD data suggest that the material is suitable for \textit{in-situ} compression testing and shape recovery due to SME.}
    \label{fig:DSC_and_XRD}
\end{figure}

\section{Results and Discussion}
\begin{wrapfigure}[14]{r}{0.5\textwidth}
    \vspace*{-4em}
    \includegraphics[width=\linewidth]{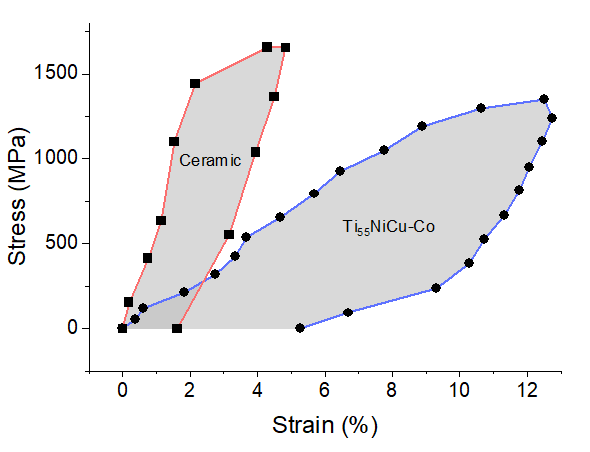}
    \caption{Stress-strain curve calculated for ceramic and metallic pillars of similar dimension.}
    \label{fig:force-displacement}
\end{wrapfigure}

Figures~\ref{fig:DSC} and \ref{fig:XRD} display the DSC and XRD characterization results. The onset and endset temperatures of the DSC transformation peaks correspond to the following transformation temperatures: $A\textsubscript{s} = \qty{244}{\celsius}$, $A\textsubscript{f} = \qty{255}{\celsius}$, $M\textsubscript{s} = \qty{-81}{\celsius}$, and $M\textsubscript{f} = \qty{-84}{\celsius}$. XRD analysis confirmed a high tetragonal phase fraction at RT before liquid nitrogen treatment and a predominantly monoclinic phase afterward, validating the DSC results and enabling shape memory testing.

Figure~\ref{fig:force-displacement} presents the stress-strain curves of cylindrical pillars subjected to \textit{in-situ} compression experiments, conducted in an SEM under vacuum conditions. While the ceramic material was compressed up to \qty{4}{\percent} strain, at which point the stress reached about \qty{1650}{\MPa}, the metallic sample experienced a maximum compression by more than \qty{12}{\percent} at a maximum stress of $\sim\qty{1350}{\MPa}$. As expected, a much higher Young's-modulus is apparent in the initial slope of the ceramic compared to the metallic material. Lai et al. \cite{Lai2013} observed a similar stress-strain hysteresis in their study on superelastic CeO$_2$-ZrO$_2$ pillars under compression, but at higher stresses. In contrast, our material exhibited an overall lower stress level, indicating more compatible lattice during transformation.
\begin{figure}[t]
    \centering
    \includegraphics[width=1.0\linewidth]{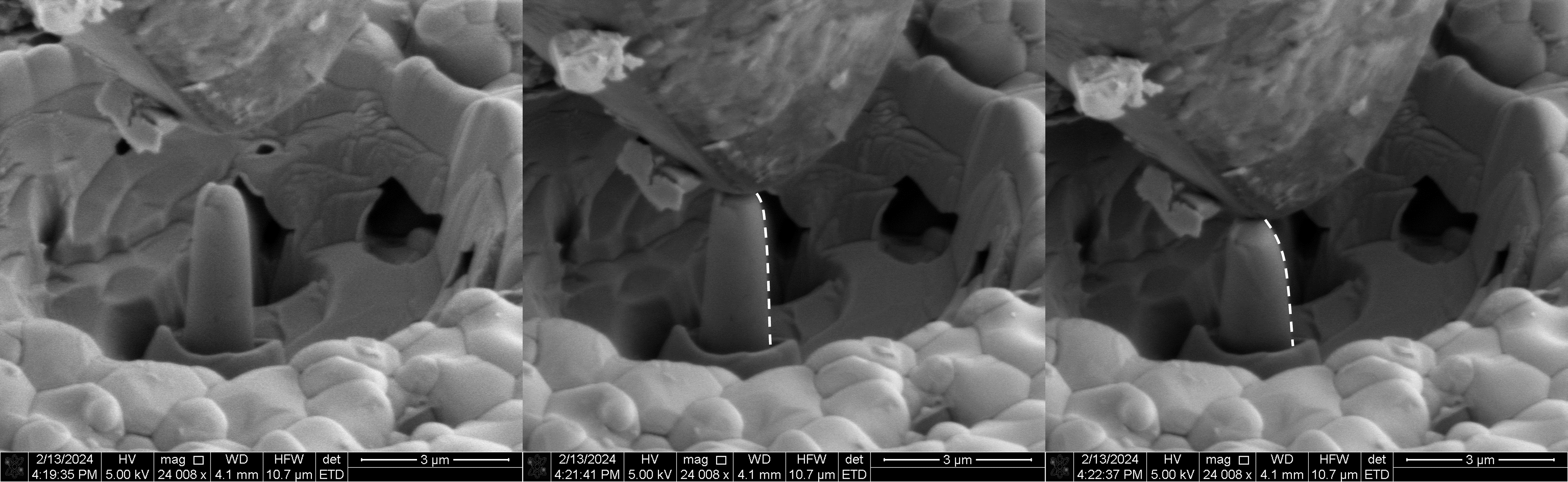}
    \caption{Deformation of the micro-pillar (left) during the compression experiment. The shape of the pillar silhouette is highlighted by a dashed line, indicating a permanent deformation between the start (middle) and end (right) of the compression test.}
    \label{fig:deformation}
\end{figure}
The curves display the characteristic behavior of shape memory materials to a degree, with an initial elastic response followed by a plateau region, where deformation occurs via twin boundary motion at nearly constant stress. Notably, the overall shape of the curve for a pillar made of Ti$_{55}$NiCu-Co appears very similar to that of the ceramic material, indicating comparable deformation characteristics.

It is important to note that the stress-strain values presented should be interpreted with caution, as they were derived from SEM images using software that tracks relative movements, rather than from actual force measurements. Consequently, there is a significant potential for drift and error in these values.

The deformation of a ceramic pillar is captured in the SEM images in Figure~\ref{fig:deformation}, where a dashed line marks the evolving silhouette. The deformation progressed until the pillar leaned substantially, approaching structural failure at its base while maintaining contact with the bulk material.

\begin{figure}[ht]
    \centering
    \includegraphics[width=1.0\linewidth]{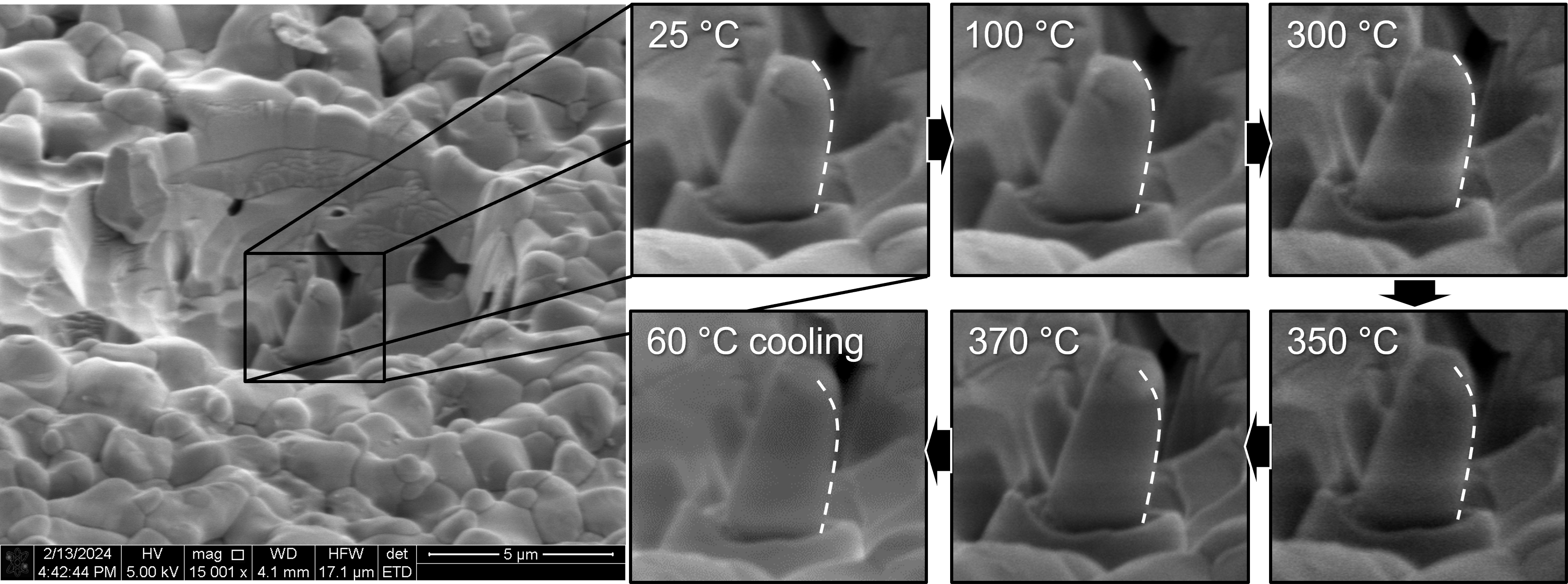}
    \caption{Shape recovery of the micropillar from the overview on the left. Part of the silhouette at \qty{25}{\celsius} is overlaid for sample temperatures of \qtylist{100;300;350;370;60}{\celsius} to indicate the shape recovery.}
    \label{fig:shape_recovery}
\end{figure}

After removing the indenter tip from the viewing area, a controlled heating process was initiated. Images were captured at various temperatures to track the pillar’s morphological evolution (Figure~\ref{fig:shape_recovery}). Up to \qty{300}{\celsius}, no significant changes were observed. However, at \qty{350}{\celsius} and subsequently at \qty{370}{\celsius}, the pillar visibly increased in height, indicating partial shape recovery. This change was attributed to the transformation from detwinned monoclinic martensite to the more symmetric tetragonal austenite phase. After cooling to \qty{60}{\celsius}, the recovered shape remained stable.
Further analysis suggests that optimizing the middle eigenvalue $\lambda_2$ and transformation strain compatibility can enhance the reversibility of shape recovery. Future work will explore doping strategies to fine-tune the transformation temperatures and improve the mechanical response of Zirconia-based shape memory ceramics.


\section{Conclusion}
Our study demonstrates the feasibility of \textit{in-situ} observation of shape recovery in a Zirconia-based ceramic material using the proposed experimental setup. By manipulating the crystallographic lattice and employing advanced microscopy techniques, we observed promising results in inducing reversible shape changes in ceramic micropillars. Further investigations are warranted to elucidate the underlying mechanisms and optimize transformation temperatures for achieving SME and superelasticity in single-grain pillar structures. This research paves the way for the development of advanced ceramic materials with tailored mechanical properties, potentially impacting a wide range of applications.

\printbibliography

\end{document}